\documentclass{aa}  

\usepackage{graphicx}
\usepackage{txfonts}
\usepackage{natbib}

\usepackage{aalongtable,lscape}
\usepackage{longtable}

\begin{document}

\title{CNO and F abundances in the globular cluster M22 (NGC~6656)} 

\titlerunning{CNO, Na, F and Fe in M22}
\authorrunning{Alves-Brito et al.}

\author{
A.~Alves-Brito\inst{1,2} \and
D.~Yong\inst{2}\and
J.~Mel\'{e}ndez\inst{3} \and
S.~V\'asquez\inst{1} \and
A.~I.~Karakas\inst{2} 
}

\institute{
Departamento de Astronom\'{i}a y Astrof\'{i}sica, 
Pontificia Universidad Cat\'{o}lica de Chile,
Av. Vicu\~{n}a Mackenna 4860, Macul, 782-0436, Santiago, Chile;
 \and
Research School of Astronomy and Astrophysics,
The Australian National University, Cotter Road, Weston, ACT 2611, Australia;
\email{abrito@mso.anu.edu.au} 
\and
Departamento de Astronomia, IAG, Universidade de S\~{a}o Paulo, Rua do Mat\~{a}o 1226, 
Cidade Universit\'{a}ria, S\~{a}o Paulo 05508-900, Brazil;
}

\date{Received: ; accepted: }

\abstract
{Recent studies have confirmed the long standing suspicion that M22 shares a
metallicity spread and complex chemical enrichment history similar to that observed
in $\omega$ Cen. M22 is among the most massive Galactic globular
clusters and its colour-magnitude diagram and chemical abundances reveal the
existence of sub-populations.
}
{To further constrain the chemical diversity of M22, necessary to interpret its
nucleosynthetic history, we seek to measure relative abundance ratios of key
elements (carbon, nitrogen, oxygen, and fluorine) best studied, or only available,
using high-resolution spectra at infrared wavelengths. 
}
{High-resolution (R = 50,000) and high S/N infrared spectra were acquired of nine red
giant stars with Phoenix at the Gemini-South telescope. Chemical
abundances were
calculated through a standard 1D local thermodynamic
equilibrium analysis using Kurucz model atmospheres.
}
{We derive [Fe/H] = $-$1.87 to $-$1.44, confirming at infrared
wavelengths that M22 does present a [Fe/H] spread. We also find large C and N
abundance spreads, which confirm previous results in the literature but based
on a smaller sample. 
Our results show a spread in A(C+N+O) of $\sim$ 0.7 dex.
Similar to mono-metallic GCs, M22 presents a strong [Na/Fe]-[O/Fe]
anticorrelation as derived from Na and CO lines in the K band. For the first
time we recover F abundances in M22 and find that it exhibits a 0.6 dex
variation. 
We find tentative evidence for a flatter A(F)-A(O) relation compared to higher
metallicity GCs. 
}
{Our study confirms and expands upon the chemical diversity seen in this complex
stellar system. All elements studied to date show large abundance spreads which 
require contributions from both massive and low mass stars.}


\maketitle
%

\section{Introduction}

The idea that globular clusters (GCs) are the simplest example of a 
stellar population \citep{rb86}, 
that is, all their stars having the same age and chemical
composition, has been challenged over the years.
On the one hand, detailed chemical abundance analyses have
shown that almost all GCs present negligible internal abundance variations of
iron-peak elements, whereas an unexpected large star-to-star abundance
variation is found for light elements (Li, C, N, O, Na, Mg and Al), within
which a strong anticorrelation is present in the O:Na and Mg:Al pairs. As these
anomalies are found in both giant stars and turnoff and sub-giant
stars \citep[see e.g.][]{cannon98}, the most plausible explanation is that they are 
primordial in origin, that is, they
must be present in the gas from which these stars were formed. In this scenario,
intermediate mass (3-8M$_{\odot}$) asymptotic giant branch (AGB) and massive
stars acting in the early stages of
globular cluster formation are the strongest polluter candidates
\citep[see][for an extensive review]{gratton04}. 
Additionally, very deep photometry of some of the most massive 
Galactic globular clusters has revealed color-magnitude diagrams (CMDs) 
that display multiple stellar populations \citep[see e.g.][and references therein]{piotto09} and such behavior has not
yet been fully understood \citep[see, e.g.,][for a recent review and alternative
scenario]{vc11}. 

In this context NGC~6656 (M22, l = 9.89$^{\rm o}$, b = $-$7.55$^{\rm o}$), an
inner halo GC located at R$_{\sun}$ = 3.2 kpc, is one of the most interesting globular clusters
for detailed abundance analysis because previous photometric
\citep[e.g.][]{hh79} and spectroscopic \citep[e.g.][]{nf83}
studies revealed a chemical heterogeneity for this
cluster. Photometrically, \citet{piotto09} identified multiple stellar
populations on the sub-giant branch of M22. 
Detailed abundance analyses based on a few stars
and using high-resolution CCD spectra, albeit
limited to a spectral resolution of R $\sim$ 20,000
\citep[see, e.g.,][for more details]{gratton82,pila82,brown90,bw92} found conflicting results
regarding the abundance spread in M22. Studying six giant stars
in M22, \citet{pila82} found a large spread of 0.5 dex in metallicity for this
GC. These findings were also supported by \citet{l91} who studied 10 giant stars
and found a variation of $\Delta$[Fe/H] = +0.3 dex. 
On the other hand, others found no evidence for a metallicity dispersion
\citep[e.g.][]{co81,gratton82,i04}. More recently, a coherent picture is
developing which confirms a metallicity spread in M22 
using optical spectra with a range in resolution 
\cite[R = 2,000-60,000;][]{marino09,marino11,dacosta09}.
Thus, photometric and spectroscopic studies suggest that M22 shares
striking similarities to $\omega$ Cen --- the most
massive Galactic globular cluster that shows a wide range of chemical
abundance variations, as it was first proposed by \citet{hess77}. 

CNO group abundances are especially important for 
understanding the properties of multiple population globular clusters
\citep{cassisi08}. 
For M22, CNO group abundances have been reported in the literature for
a few stars.
Using photometry, \citet{frogel83} reported a large spread in CO within M22.
Spectroscopically, \citet{nf83} found a chemical inhomogeneity in M22
based on the behaviour of the CH, CN and A(Ca) spectral indices for a large number of giant
stars \citep[see also][for a larger sample]{at95}. 
In particular, they found a direct correlation between the variation of CN
and the Ca~II H and K lines. \citet{brown90} studied seven stars in M22 and
found a large range of N abundances in their sample that was difficult
to explain by CNO cycling and mixing. \citet{kayser08} compared the relative CN
and CH line strengths of 97 stars and concluded that the number of CN-weak
stars in M22 is higher than that of the CN-strong. More recently
\citet{marino11} studied 35 red giants in M22 using high-resolution optical
spectra and found C-N anticorrelations in the subsample analysed.

Fluorine is a relatively poorly studied element that may provide 
key insights into globular cluster chemical evolution. 
\citet{jorissen92} were the first to determine F abundances outside the solar
system. Almost two decades later, it has been known that F can be
synthesized in AGB stars, and it is produced in the He-intershell 
along with C and $s$-process elements \citep{jorissen92,lugaro04}; however,
the nucleosynthetic origin of fluorine in the Galaxy is still a mystery (see,
e.g., the Introduction of \citet{ab11} for a recent summary). 
Despite its importance as a tracer of AGB nucleosynthesis, F has been only
measured in three Galactic globular clusters to date --- 
$\omega$ Cen \citep{cunha03}, M4 \citep{smith05} and NGC~6712 \citep{yong08}. 

In this work we present the first abundance estimates of C, N, O, F, Na and Fe
for giant stars in M22  using spectra taken at infrared
wavelengths through the high-resolution, near-infrared Phoenix spectrograph
at Gemini-South.
We investigate the CNO group and Fe abundance variations in M22 and the
behavior of F with respect to O and also to the $s$-process elements. 
These data provide additional critical constraints upon the role of massive and
low mass stars
in the chemical evolution of M22. Finally, we compare our results
with chemical abundance patterns in globular clusters and field stars in order
to provide valuable information on how the different Galactic populations were
formed and have evolved. 

The paper is organized as following. Section 2 describes the observations.
Section 3 presents the data analysis. The results and discussions are outlined
in Sect. 4. Finally, Sect. 5 summarizes the main conclusions.

\section{Observations}
\label{s:observations}

\begin{figure}
   \centering
   \includegraphics[width=8.5cm]{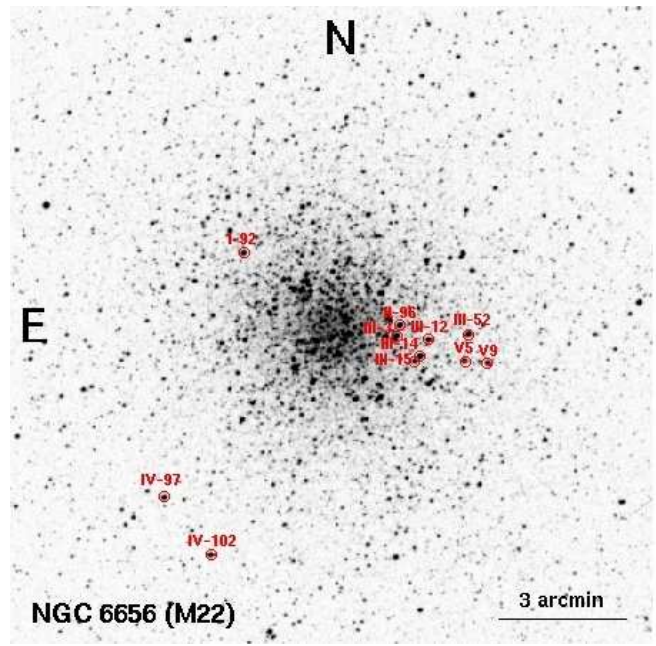}
   \caption{The finding chart of M22 from the optical Digitized Sky Survey. 
   The targets are clearly marked following the nomenclature
   defined in \citet{pc94}. The field's orientation is shown and the displayed
   sky area is of 15 x 15 arcmin$^{2}$. The coordinates (J2000) of the image centre are
   $\alpha$ = 18$^{\rm h}$36$^{\rm m}$24$^{\rm s}$.20, $\delta$ = $-$23$^{\rm
   o}$54'12$^{\rm "}$.20}
              \label{f:finding}%
    \end{figure}

In this work we take advantage
of the high-resolution infrared spectrograph Phoenix at Gemini-South
\citep{hinkle03} to carry out a detailed abundance analysis of bright giant stars in M22. The
targets were selected based on the study of proper motions and radial
velocities presented by \cite{pc94} and, additionally, by
analysing the CMD from the Two Micron All Sky Survey \citep[2MASS,][]{s06}.
We selected 11 bright
giant stars that could be observed with reasonable exposure-times, 6.93 $\leq$
H$_{2MASS}$ $\leq$ 7.97 mag, with high membership probabilities (P $>$ 87\%;
see Fig. \ref{f:finding}). In addition, for a more
reliable comparison with results previously presented for
this cluster, we included in our sample four out of six stars studied by
\citet{pila82}, who found a spread of 0.5 dex in metallicity for
these four stars.

The stars were observed in service mode between 2009 March-July.
The slit width-size of 0.34 arcseconds was
used in order to reach a resolution of R = 50,000.
To study the sample, our observing programme
was executed at two different filters, H6420 and K4308, which cover a
wavelength range from 1552 to 1558 nm (H band) and from 2330 to 2340 nm (K
band). 
Each target was
observed at two different positions along the slit and
the exposure times ranged from 316 to 1220 s. In addition to standard
calibrations (10 bias and 10 dark exposures), 
in each night of our observing run we included spectra of rapidly rotating hot
stars in order to correct the spectra for fringing (H band) and telluric lines
(K band).
The nominal signal-to-noise ratio (S/N) per pixel in the H band ranges from 60-140,
while in the K band it varies from 130-160. 
In Table \ref{t:program} we list the stars, along with their magnitudes, date of
observations and S/N estimates in both bands.
The spectra were reduced with IRAF\footnote{IRAF is distributed by the National
Optical Astronomy Observatory, which is operated by the Association of
Universities for Research in Astronomy (AURA) under cooperative agreement with
the National Science Foundation.} employing standard procedures as described
in details in \citet{smith02} and \cite{melendez03}, which include dark
and sky subtraction, flatfield correction, spectrum extraction, 
wavelength calibration and telluric correction.

As noted in \citet{ab11}, 
the CNO group elements are better studied in the infrared, 
where their lines are more numerous and the continuum is formed deepest in the 
layers of the stellar atmosphere due to the opacity minimum of H$^{-}$ near the H
band. 
For red giant stars, F measurements are only available from the HF molecule in the K band.

\section{Data analysis}
\label{s:analysis}

\subsection{Atmospheric parameters}

\subsubsection{Photometry}

\begin{figure}
\centering
 \includegraphics[width=8.5cm]{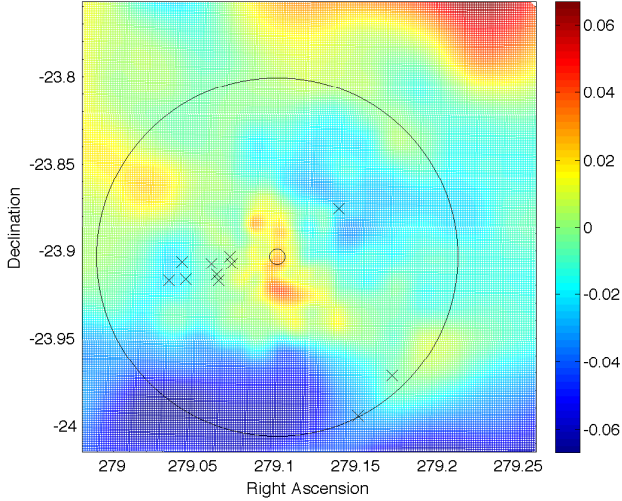}
   \caption{Extinction map for M22 in equatorial coordinates (J2000).
   The {\it small} and {\it large} circles indicate, respectively, the GC's
   centre and the area where the program stars ({\it crosses}) lie in the GC's field. The colour
   bar gives the excesses E(B-V) for the extinction map. 
   The zero point
   is arbitrary and it is calculated by comparison with isochrones \citep[see][for more details on the techniques and methods used
to create this figure]{ag11,ag11b}. The orientation is different from Fig.
1 (North-East points to the upper right corner). }
              \label{f:map_extinct}%
    \end{figure}

Due to its relatively low galactic latitude, M22 presents a high mean
interstellar extinction, E$(B-V)$ = 0.34 \citep{harris96}. 
In addition, as M22 is projected relatively near the Galactic
centre (l = 9.89$^{\rm o}$, b = $-$7.55$^{\rm o}$), the extent of its differential
extinction has been a controversial topic. 
Using the Balmer strengths of blue
horizontal branch stars, \citet{cro88} found a $\Delta$$E(B-V)$ less
than 0.08 mag for M22. Similar values were also found by \citet{mini92} using
polarization. Based on wide-field B, V and I photometry, \citet{monaco04} found 
$\Delta$$E(B-V)$ $\sim$ 0.06 mag. They argue
that the likely multiple stellar populations in this cluster is just an
artifact created by the reddening variations across the area of the GC.
Alternatively, we show in
Fig. \ref{f:map_extinct} the extinction map for M22 that was kindly provided
by Dr. Javier Alonso-Garcia. As can be seen, the map suggests that the 
total range in $E(B-V)$ is small, less than 0.06 mag within approximately 15 arcmin
from the GC's centre. The differential extinction is particularly important because it adversely
impacts upon the assignment of star's membership probabilities and
also increases the uncertainties on the temperatures and surface gravities
obtained photometrically in reddened GCs.

To obtain the photometric effective temperatures, we used the magnitudes listed in
Table \ref{t:program} altogether with the empirical temperature scale of \citet{alonso99}.
The 2MASS  magnitudes were transformed to the
Telescopio Carlos S\'anchez (TCS) system using relations by \citet{carpenter01} and
\citet{alonso98}. The colours were dereddened by adopting a fixed reddening E(B-V) = 0.34
\citep{harris96}
and reddening ratios E(V-K)/E(B-V) = 2.744 and E(J-K)/E(B-V) =
0.52 \citep{rl85}. However, as the
(V-K)$_{\rm 0}$ colours provide more accurate temperatures due to its long
baseline in wavelength, we have chosen to adopt the T$_{\rm V-K}$ as the
final photometric temperatures for our program stars. 
Alonso's calibrations are metallicity dependent, and thus we have adopted a
metallicity of [Fe/H]\footnote{In this work we used the standard spectroscopic notation A(X) = log[n(X)/n(H)]
+12 and [X/Y] = log[n(X)/n(Y)]$_{*}$ - log[n(X)/n(Y)]$_{\odot}$} = $-$1.64 for M22 \citep{harris96}. 

Surface gravities ($\log g$) were derived with the classical relation,
adopting T$_{\odot}$ = 5780 K, Mbol$_{\odot}$ = 4.75, $\log g_{\odot}$ =
4.44 dex, and M$_{*}$ = 0.80M$_{\odot}$. In addition, we assumed a distance
modulus of (m-M)$_{V}$  = 13.60, and bolometric corrections BC$_{V}$ as given
by \citet{alonso99}.

\begin{table*}
\begin{flushleft}
\caption{Program stars}
\label{t:program}      
\centering          
\begin{tabular}{lclllllllll}     
\noalign{\smallskip}
\hline\hline    
\noalign{\smallskip}
\noalign{\vskip 0.1cm} 
Star$^{a}$ & R.A.$^{b}$ & Dec.$^{b}$ & V$^{a}$ & J$^{c}$ & H$^{c}$ & K$_{c}$$^{c}$ & H$^{d}$ band & K$^{d}$ band & (S/N)$_{H}$$^{d}$ & (S/N)$_{k}$$^{d}$ \\        
(1) & (2)  & (3)  & (4) & (5) & (6)  & (7)  & (8)  & (9)  & (10) & (11)  \\                    
\hline 
1-22,V5 & 18 36 10.63 & -23 54 59.4 & 10.86  &   7.72 & 6.99 &   6.73 & ...  & 2009-04-14 : 4x75 s  & ... & 120 \\   
IV-97	& 18 36 41.11 & -23 58 18.3 & 11.08  &   7.75 & 6.99 &   6.75 & 2009-03-16 : 4x85 s  & 2009-04-14 : 4x81 s  & 110 & 130 \\   
IV-102  & 18 36 36.12 & -23 59 39.8 & 11.09  &   7.75 & 6.99 &   6.76 & 2009-03-16 : 4x87 s  & 2009-04-14 : 4x82 s  & 110 & 150 \\    
V9      & 18 36 08.23 & -23 55 01.6 & 11.13  &   7.70 & 6.92 &   6.67 & 2009-03-14 : 4x80 s  & ...		    & 100 & ... \\   
III-3	& 18 36 17.44 & -23 54 26.6 & 11.15  &   7.82 & 7.01 &   6.78 & 2009-03-16 : 4x87 s  & 2009-04-14 : 4x83 s  & 100 & 160 \\    
III-14  & 18 36 15.25 & -23 54 49.9 & 11.18  &   7.78 & 6.99 &   6.74 & 2009-06-01 : 4x97 s  & 2009-07-16 : 4x139 s &  60 & 150 \\   
III-15  & 18 36 15.55 & -23 55 01.6 & 11.30  &   8.11 & 7.34 &   7.13 & 2009-03-17 : 4x119 s & 2009-04-14 : 4x115 s & 110 & 130 \\   
III-12  & 18 36 14.45 & -23 54 26.8 & 11.48  &   8.38 & 7.61 &   7.40 & 2009-06-24 : 4x225 s & 2009-07-16 : 4x257 s & 130 & 150 \\   
III-52  & 18 36 10.12 & -23 54 22.2 & 11.50  &   8.42 & 7.64 &   7.45 & 2009-03-17 : 4x195 s & 2009-04-14 : 4x194 s & 140 & 140 \\   
I-92	& 18 36 33.22 & -23 52 32.8 & 11.51  &   8.58 & 7.82 &   7.65 & 2009-04-15 : 4x232 s & 2009-04-14 : 4x232 s & 120 & 140 \\	
II-96	& 18 36 17.18 & -23 54 11.6 & 11.60  &   8.66 & 7.96 &   7.78 & 2009-04-15 : 4x305 s & ... & 150 & ...  \\   
\hline                  
\end{tabular}
\begin{minipage}{.88\hsize}
 Notes.--- (a): Names and V magnitudes are taken from
 \citet{pc94}; (b): units of right ascension (J2000) and declination (J2000) are, respectively,
 hours, minutes, seconds, degrees, arcminutes, arcseconds; (c): J, H and K$_{s}$ are 2MASS magnitudes; (d): dates, integration times and S/N for the
 spectroscopic observations in the two different bands used.\\
\end{minipage}			
\end{flushleft}
\end{table*}  

\subsubsection{Spectroscopy}

All program stars were recently studied by \citet{marino11} using high-resolution
optical spectra (37,500 $\leq$ R $\leq$ 60,000). 
Hence, due to both the
controversial issues regarding the derivation of the photometric atmospheric
parameters and also to the fact that there are no numerous iron lines at
infrared wavelengths, we adopted the spectroscopic atmospheric parameters
derived by \citet{marino11} using numerous Fe lines. However, we have used our
own infrared data to constrain the final metallicities, which were iteratively
adjusted along with the CNO abundances. 

Final atmospheric parameters and kinematics are
presented in Table \ref{t:atmos}. 
Interestingly, examination of Table \ref{t:atmos} reveals that the
photometric parameters, which were obtained by assuming a uniform reddening of
$E(B-V)$ = 0.34 mag, are in good agreement with the spectroscopic ones. The mean
difference (photometry - spectroscopy) is --46 $\pm$ 19 ($\sigma$ = 56) K
for T$_{\rm eff}$ and +0.08 $\pm$ 0.06 ($\sigma$ = 0.20) dex for $\log g$.
Because the spectroscopic analysis is reddening free, these results indicate
that there is no significant reddening variation for the stars studied. We
reiterate that Fig. 2 also suggests that there is no significant
differential reddening for the program stars.

We note, however, that the comparison of the stellar spectroscopic parameters 
adopted in this work and previous literature values 
\citep[see,
e.g.][]{pila82,gratton82,wall87,ss89,brown90}
reveals some discrepancies.
These differences vary significantly, for a given star, 
depending on which literature study is adopted for comparison. 
We
believe that such discrepancies are not only due to the different parameters
used (e.g., reddening, distance modulus, colours) but also as a
consequence of the different techniques of analysis employed by
the different authors.
For example, for the star III-3, \citet{brown90} obtained T$_{\rm eff}$ = 4500
K, $\log g$ = 0.7 dex and $\xi$ = 2.3 km s$^{-1}$, while \citet{pila82}
found T$_{\rm
eff}$ = 4100 K, the same $\log g$ and a higher microturbulent velocity, $\xi$ = 3.5
km~s$^{-1}$.

\begin{table*}
\caption{Stellar parameters, radial velocities, and membership probabilities.}
\label{t:atmos}      
\centering          
\begin{tabular}{lcllllllllllllll}     
\noalign{\smallskip}
\hline\hline    
\noalign{\smallskip}
\noalign{\vskip 0.1cm} 
    & \multicolumn{2}{c}{Photometry$^{a}$} & &  \multicolumn{5}{c}{Spectroscopy$^{b}$}  &  &  & &  & \\
\cline{2-3} \cline{5-9} \\ 
Star & T$_{\rm eff}$ [K] & log g [dex] && T$_{\rm eff}$ [K] & $\log g$ [dex] & $\xi$ [kms$^{-1}$] & [Fe/H]$_{optical}$ & [Fe/H]$_{IR}$ & v$_{r}^{H}$ [kms$^{-1}$] &
v$_{r}^{K}$ [kms$^{-1}$] & P$^{c}$\\  
     
(1) & (2)  & (3)  && (4) & (5) & (6)  & (7) & (8) & (9)  & (10) & (11)\\                    

\hline 
IV-97	& 3974 & 0.42 && 4000 & 0.05 & 2.00 & -1.94 & -1.84 & -149.4 (0.5) & -148.2 (0.3) & 99 \\   
IV-102  & 3974 & 0.43 && 4020 & 0.20 & 2.20 & -1.97 & -1.87 & -149.1 (0.5) & -149.4 (0.4) & 99 \\    
III-3   & 3952 & 0.43 && 4000 & 0.30 & 2.20 & -1.72 & -1.62 & -148.6 (0.5) & -148.7 (0.5) & 87 \\   
III-14  & 3919 & 0.41 && 4030 & 0.35 & 2.15 & -1.82 & -1.64 & -150.2 (0.6) & -149.3 (0.3) & 99 \\      
III-15  & 4055 & 0.59 && 4070 & 0.40 & 1.85 & -1.82 & -1.72 & -148.3 (0.6) & -148.6 (0.7) & 99 \\      
III-12  & 4102 & 0.70 && 4185 & 1.00 & 1.95 & -1.65 & -1.44 & -149.1 (0.4) & -149.4 (0.4) & 98 \\      
III-52  & 4124 & 0.72 && 4075 & 0.60 & 1.75 & -1.63 & -1.54 & -148.8 (0.5) & -148.3 (0.6) & 97 \\      
I-92	& 4239 & 0.81 && 4240 & 0.75 & 1.55 & -1.75 & -1.65 & -148.9 (0.7) & -149.8 (0.3) & 88 \\       
II-96	& 4268 & 0.87 && 4400 & 1.00 & 2.10 & -1.82 & -1.60 & -148.1 (0.6) &  ...      & 99\\   
 
\hline                  
\end{tabular}
\begin{minipage}{.88\hsize}
 Notes.--- (a): Photometric stellar parameters; (b) T$_{\rm eff}$, $\log g$, $\xi$, and [Fe/H]$_{optical}$ are from \citet{marino11}, while the [Fe/H]$_{IR}$ values are obtained
 in this work (refer to the text); (c): membership probabilities from Peterson \& Cudworth (1994)\\
\end{minipage}			
\end{table*}

\subsection{Spectral synthesis}

To measure the chemical compositions of C, N, O, F, Na and Fe in the 
H band at 1555 nm (OH, CO, CN and Fe) and in the K band at 2330 nm (CO, HF and Na)
we used the same line list as used in previous studies dedicated
to the analysis of giant stars in the infrared \citep[see, e.g.][for more
details]{melendez03,melendez08,yong08}.
The HF (1-0) R9 line at 2335 nm presents
$\chi$ = 0.480 eV and $log gf$ = --3.955 dex. We have assumed isotopic ratios
$^{12}$C/$^{13}$C = 5 for all stars.

\begin{figure*}
   \centering
   \includegraphics[width=10.5cm]{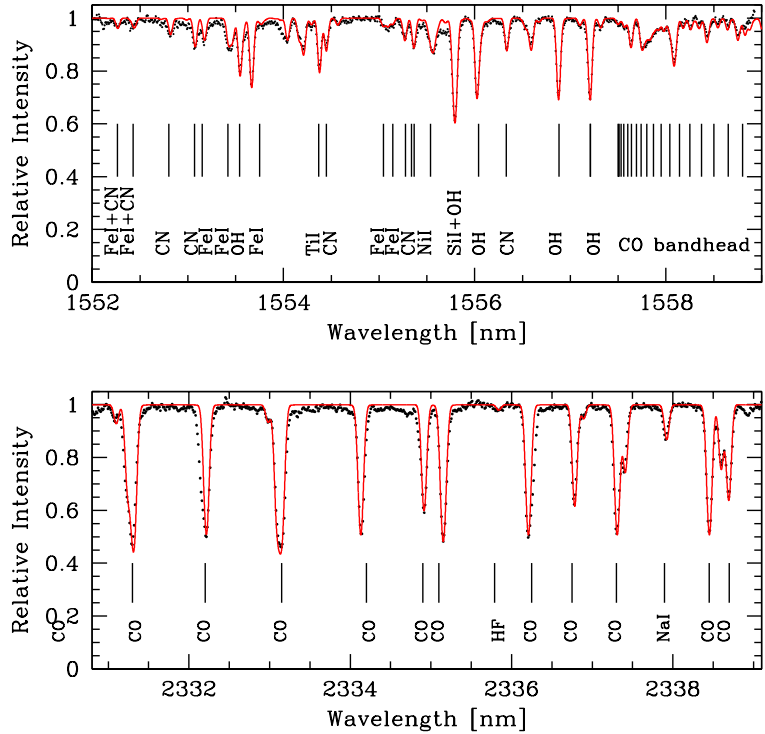}
   \caption{Observed ({\it
   points}) and best synthetic ({\it lines}) spectra of M22 III-52 in the H ({\it
   top}) and K ({\it botton}) bands. Adopted abundances can be found in Tables \ref{t:atmos} and
   \ref{t:abund}.
    Several atomic and molecular lines are identified.
   }
              \label{f:fit0}%
    \end{figure*}

\begin{figure}
   \centering
   \includegraphics[width=8.5cm]{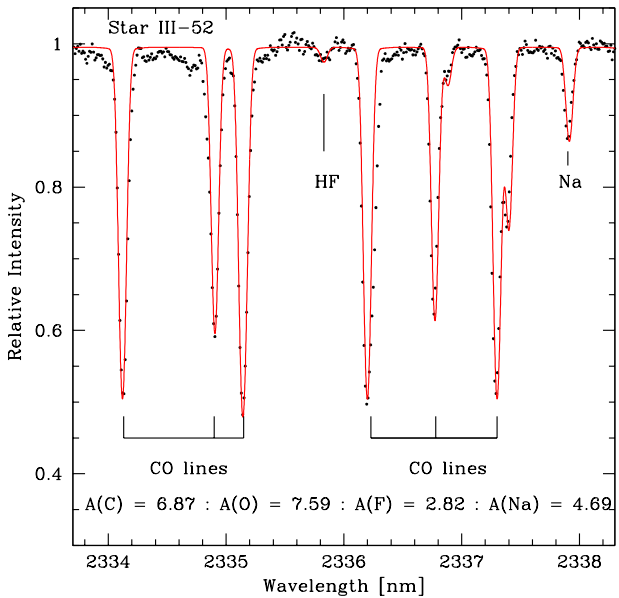}
   \caption{Example of spectral synthesis in M22 III-52 in the K band showing,
   in more details, the abundance fit for CO, F and Na lines. Atomic and
   molecular lines as well as the derived abundances are labeled in the figure.
   }
              \label{f:fit1}%
    \end{figure}

\begin{figure}
   \centering
   \includegraphics[width=8.5cm]{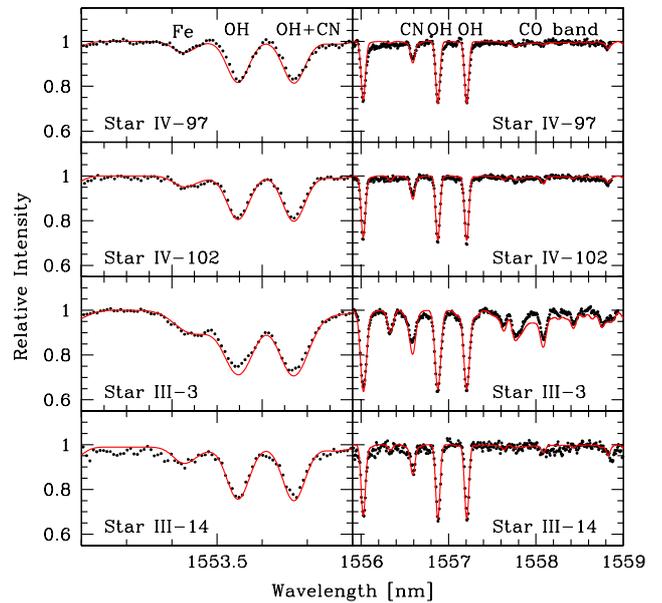}
   \caption{Example of spectral synthesis in M22: observations ({\it black
   points} and synthetic spectra ({\it red line}) for the stars IV-97, IV-102,
   III-3, and III-14 around 1553.5 nm ({\it left panel}) and 1557.5 nm ({\it right
   panel}). Adopted abundances can be found in Tables \ref{t:atmos} and
   \ref{t:abund}. Some atomic and molecular lines are also indicated in the
   figure.}
              \label{f:fit2}%
    \end{figure}

\begin{figure}
   \centering
   \includegraphics[width=8.5cm]{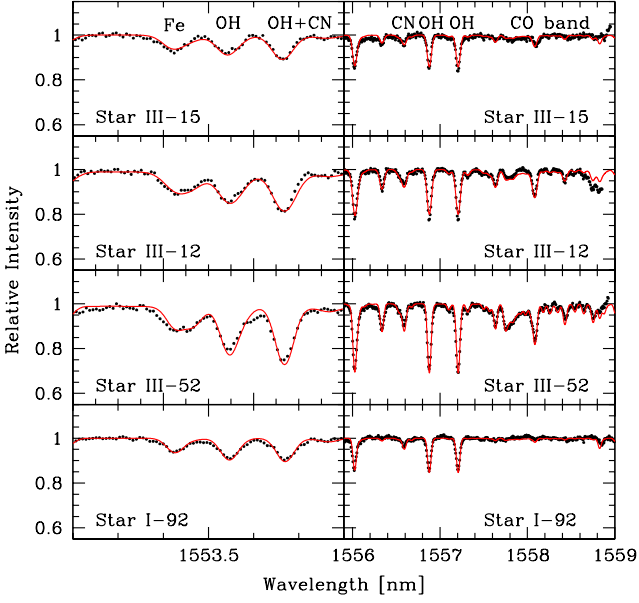}
   \caption{Same as Fig. \ref{f:fit2} but for the stars III-15, III-12,
   III-52 and I-92.}
              \label{f:fit3}%
    \end{figure}

\begin{figure}
   \centering
   \includegraphics[width=8.5cm]{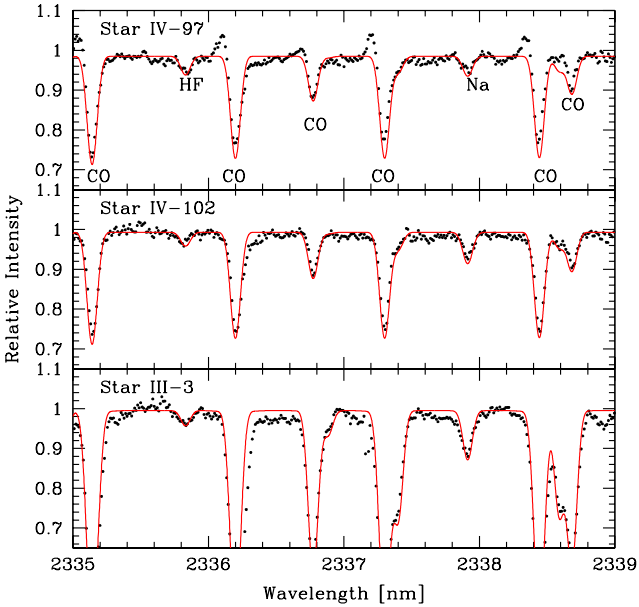}
   \caption{Example of spectral synthesis for the stars IV-97 ({\it
   top}), IV-102 ({\it middle}), and III-3 ({\it bottom}) covering the HF, CO, and Na lines in the K band. Observed spectra ({\it black
   points}) are fitted by synthetic spectra ({\it red line}) whose abundances are
   given in Tables \ref{t:atmos} and \ref{t:abund}. Atomic and molecular lines
   are indicated.}
              \label{f:fit4}%
    \end{figure}

\begin{figure}
   \centering
   \includegraphics[width=8.5cm]{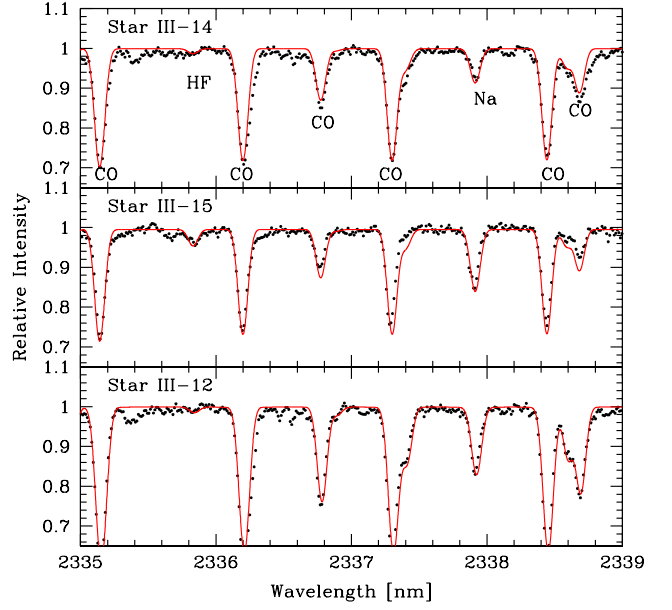}
   \caption{Same as Fig. \ref{f:fit4} but for the stars III-14, III-15
   and  III-12.}
              \label{f:fit5}%
    \end{figure}

\begin{figure}
   \centering
   \includegraphics[width=8.5cm]{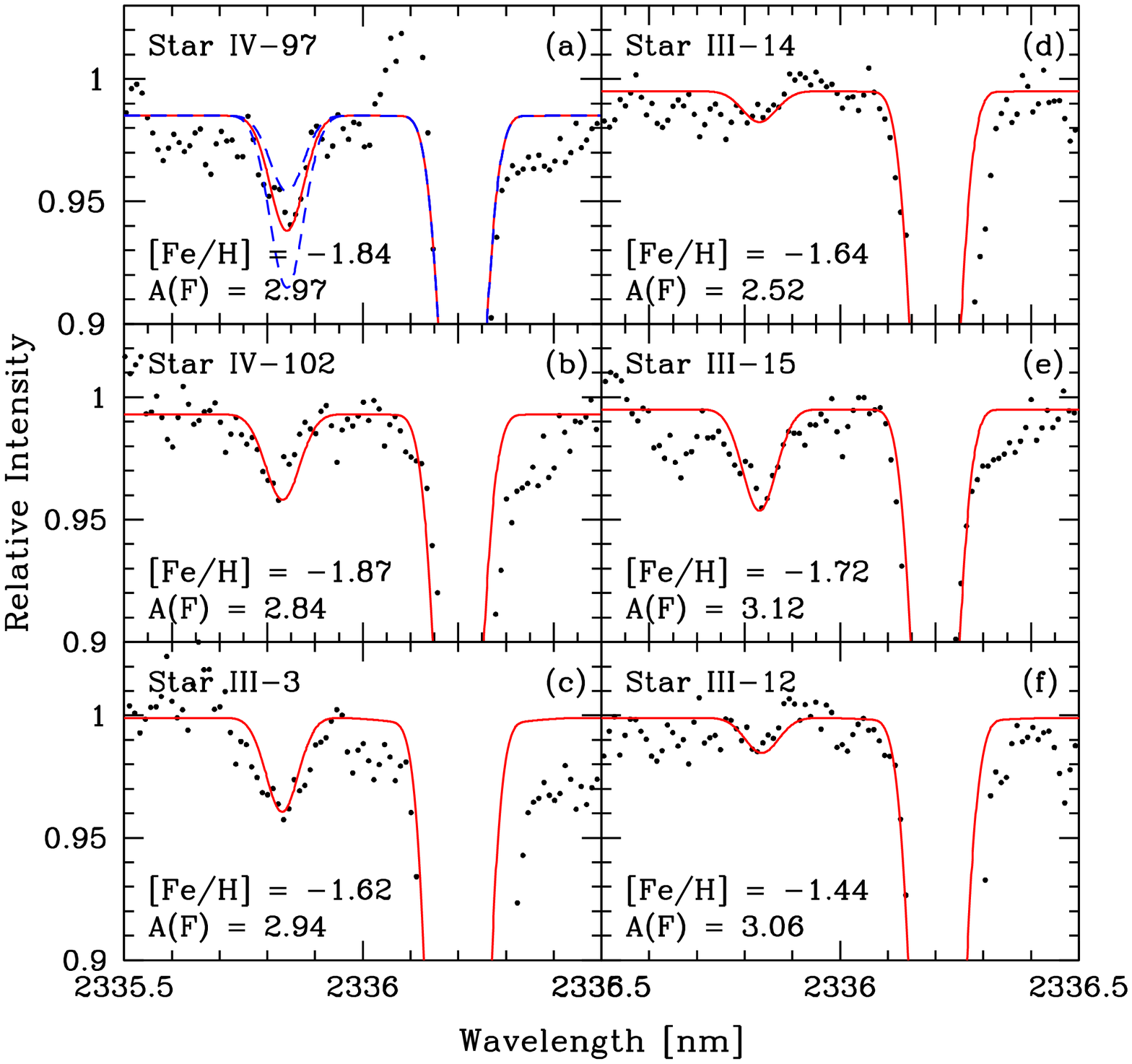}
   \caption{Comparison of observed ({\it black points}) and synthetic ({\it red line}) spectra
   for six stars in M22 around the HF molecular line. Metallicities and
   fluorine abundances are indicated. The dashed blue lines in panel (a) indicate
   variations in the best-fitting fluorine abundance by $\pm$ 0.20 dex.}
              \label{f:fit6}%
    \end{figure}

Elemental abundances were obtained using Kurucz model atmospheres
(Castelli et al. 1997) and the local thermodinamical equilibrium (LTE) spectral
synthesis code MOOG (Sneden 1973). Using the spectroscopic atmospheric
parameters given in \citet{marino11} as a first input, we reobtained the Fe
abundances and using OH lines in the H band we
obtained the [O/Fe] ratios for the sample. Carbon abundances were then obtained
by using the CO molecular bandhead at 1558 nm. Next, nitrogen abundances were
derived from CN molecular lines in the H band. The carbon abundances were
also checked using the numerous CO lines in the K band. 
Since the CNO abundances are coupled, we iterated until self-consistent
abundances were achieved. Na and F were obtained
by using the NaI line at 2337 nm and the HF molecular line at 2335 nm,
respectively. All observed spectra were convolved with Gaussian functions in
order to take the instrumental profile into account as well other broadening
effects such as macroturbulence and rotation. Except for star III-3, the
synthetic spectra in the K band, convolved with a Gaussian of typical FWHM
10 kms$^{-1}$, were slightly broader than those in the H band by approximately 2
kms$^{-1}$. Figures \ref{f:fit0}-\ref{f:fit6} illustrate the fit of synthetic
spectra to the observed ones in both H and K bands. 
  
\subsection{Error analysis}

As previously stated, the exact reddening for M22 is somewhat uncertain. 
To obtain the photometric atmospheric parameters we have adopted and applied a uniform reddening
correction of E(B-V) = 0.34 mag for all stars. Had we changed the $E(B-V)$ value
by +0.06 dex, the temperature would increase by 150 K.
Additionally, that would lead to an uncertainty of approximately 0.5 kpc in
the distance, which would translate into an uncertainty in $\log g$ of
approximately +0.20 dex. Since $\Delta$$E(B-V)$ is likely smaller than 0.06
for the cluster in general and for the program stars in particular (see Fig.
\ref{f:map_extinct}), the errors in T$_{\rm eff}$ and $\log g$ mentioned above
should be smaller. A typical uncertainty of 0.03 in $E(B-V)$ would lead to an
error of about 75 K in Teff and about 0.1 dex in $\log g$. These uncertainties
are in good agreement with those estimated by \citet{marino11} for the
spectroscopic parameters. As explained in Sect. 3.1.2, the mean difference between the
photometric and spectroscopic parameters is of --46 $\pm$ 19 ($\sigma$ = 56) K
for T$_{\rm eff}$ and +0.08 $\pm$ 0.06 ($\sigma$ = 0.20) dex for $\log g$. 

We then estimate the impact on the abundance ratios by varying the stellar
parameters of $\Delta T_{\rm eff} \approx \pm 100$\,K, 
$\Delta \log g \approx \pm 0.3$\,dex and 
$\Delta v_{\rm t} \approx 0.2$\,km\,s$^{-1}$. 
The individual values as well as the total abundance errors due to uncertainties
in the different atmospheric parameters added in quadrature are shown in
Table \ref{t:errors}.

\begin{table}
\begin{flushleft}
\caption{Sensitivities in the abundance ratios for the star III-52. 
The atmospheric parameters were changed by 
$\Delta T_{\rm eff} = \pm 100$\,K, $\Delta \log g = \pm 0.30$\,dex, and  
$\Delta v_{\rm t} = \pm 0.20$\,km\,s$^{-1}$. 
The total internal uncertainties are
given in the last column.}

\label{t:errors}      
\centering          
\begin{tabular}{lcccc}     
\noalign{\smallskip}
\hline\hline    
\noalign{\smallskip}
\noalign{\vskip 0.1cm} 
Abundance & $\Delta$T$_{\rm eff}$  & $\Delta$log g & $\Delta$v$_{\rm t}$ & ($\sum x^{2}$)$^{1/2}$ \\        
   
(1) & (2)  & (3)  & (4) & (5)   \\ 
\hline
\hbox{A(C)}   &    +0.06   &    +0.07  &	+0.03	  &  +0.10 \\
\hbox{A(N)}   &    +0.07   &    -0.10  &	+0.04	  &  +0.13 \\
\hbox{A(O)}   &    +0.09   &    +0.05  &	-0.01	  &  +0.10 \\
\hbox{A(F)}   &    +0.15   &    -0.08  &	+0.03	  &  +0.17 \\
\hbox{A(Na)}  &    +0.06   &    -0.04  &	-0.04	  &  +0.08 \\
\hbox{A(Fe)}  &    +0.02   &    +0.03  &	+0.02	  &  +0.04 \\
\hline          	   		  		     	   
\end{tabular}
\end{flushleft}
\end{table}  

\section{Results and discussions}

\subsection{Radial velocity and cluster membership}

Radial velocities were obtained using the IRAF task {\it
rvidlines} in each wavelength setting by using clean isolated lines. 
In the H band we found a mean radial velocity of v$_{\rm r}$ = $-$148.9 $\pm$ 0.2 ($\sigma$ = 0.6, 9 stars)
kms$^{-1}$, while in the K band it was v$_{\rm r}$ = $-$148.9 $\pm$ 0.2
($\sigma$ = 0.5) kms$^{-1}$. 
Additionally, our results are in excellent agreement with 
literature values where, for example, \citet{harris96} provided a mean value of 
v$_{\rm r}$ = $-$148.9 $\pm$ 0.4 kms$^{-1}$.
Stars III3, IV97 and III-52 were recently analysed by \citet{marino09} who found
mean values of 
v$_{\rm r}$ = $-$148.16, $-$149.84 and $-$153.22 kms$^{-1}$, respectively.
These comparisons are an independent check that all program stars belong
to M22.

\subsection{Iron}

\begin{table*}
\begin{flushleft}
\caption{Final abundances$^{a}$.}
\label{t:abund}      
\centering          
\begin{tabular}{lcccccccccccccc}     
\noalign{\smallskip}
\hline\hline    
\noalign{\smallskip}
\noalign{\vskip 0.1cm} 
Star & A(C) & A(N) & A(O) & A(F) & A(Na) & A(C+N) & A(C+N+O) & [C/Fe] & [N/Fe] &
[O/Fe] & [F/O] & [Na/Fe] & $<$[s/Fe]$>$$^{b}$  \\  
\noalign{\vskip 0.1cm}       
(1) & (2)  & (3) & (4) & (5) & (6) & (7)  & (8) & (9) & (10) & (11) & (12) & (13) & (14)  \\                    
\hline
\noalign{\smallskip}
\noalign{\vskip 0.1cm} 
IV-97	& 5.82 & 6.41 & 7.24 & 2.97            &  4.21  & 6.51  & 7.31  & -0.76  & 0.43 & 0.36 & -0.11 & -0.12 & -0.06 \\   
IV-102  & 5.69 & 7.18 & 7.26 & 2.84            &  4.46  & 7.19  & 7.53  & -0.86  & 1.23 & 0.41 & -0.26 &  0.16 & -0.09 \\    
III-3   & 6.79 & 7.53 & 7.86 & 2.94            & 4.66   & 7.60  & 8.05  & -0.01  & 1.33 & 0.76 & -0.76 &  0.11 &  0.26 \\   
III-14  & 5.77 & 7.51 & 7.54 & $\leq$2.52      & 4.49   & 7.52  & 7.83  & -1.01  & 1.33 & 0.46 & -0.86 & -0.04 &  0.03 \\	 
III-15  & 5.84 & 7.23 & 7.01 & 3.12            & 4.96   & 7.25  & 7.45  & -0.86  & 1.13 & 0.01 &  0.27 &  0.51 &  0.00 \\	 
III-12  & 6.71 & 7.60 & 7.68 & $\leq$3.06      & 5.08   & 7.65  & 7.97  & -0.27  & 1.22 & 0.40 & -0.46 &  0.35 &  0.37 \\	 
III-52  & 6.87 & 7.21 & 7.59 & 2.82            & 4.69   & 7.37  & 7.79  & -0.01  & 0.93 & 0.41 & -0.61 &  0.06 &  0.34 \\	 
I-92	& 5.91 & 6.90 & 7.33 & ...             & 4.29   & 6.94  & 7.48  & -0.86  & 0.73 & 0.26 &  ...  & -0.23 & -0.06  \\    
II-96	& 5.96 & 6.85 & 7.63 & ...             &  ...   & 6.90  & 7.71  & -0.86  & 0.63 & 0.51 &  ...  &   ... & -0.03 \\	
\hline                  
\end{tabular}
\begin{minipage}{.88\hsize}
 Notes.--- (a): Calculated by using the spectroscopic T$_{\rm eff}$, $\log g$, $\xi$ and [Fe/H]$_{\rm IR}$ given in Table \ref{t:atmos}. We have adopted A(C,N,O)$_{\odot}$ =
 8.42, 7.82, and 8.72; A(F,Na,Fe)$_{\odot}$ = 4.56, 6.17 and 7.50. (b):
 $<$[s/Fe]$>$ stands for ([Ba/Fe] + [La/Fe])/2, whose abundances were taken from \citet{marino11}.
\end{minipage}			
\end{flushleft}
\end{table*}

\begin{figure}
   \centering
   \includegraphics[width=8.5cm]{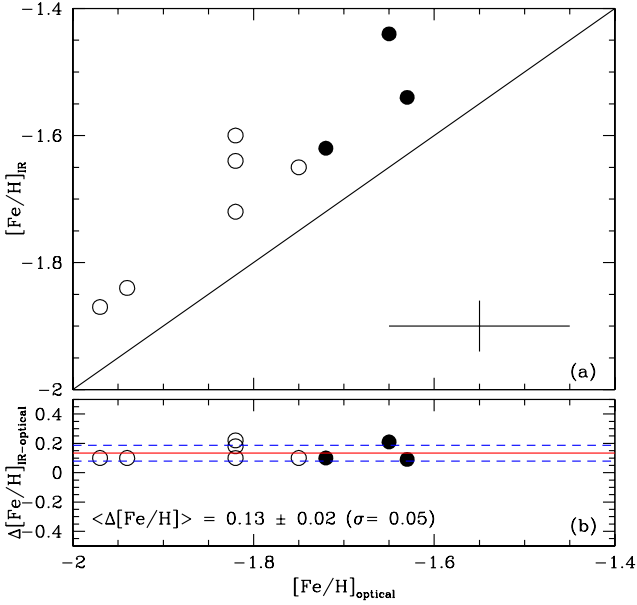}
   \caption{(a): Comparison of spectroscopic metallicities obtained in the infrared (this
   work) versus in the optical \citep{marino09} for $s$-poor ({\it open
   circles}) and $s$-rich ({\it filled circles}) stars. Typical uncertainties are
   shown. (b): [Fe/H] residuals (ours minus optical values), where the dashed
   lines represent 1$\sigma$ scatter over the mean value found ({\it full
   line}).}
              \label{f:fe_comp}%
    \end{figure}

The program
stars have metallicities ranging from [Fe/H] = $-$1.87 to $-$1.44, which
represent a metallicity
variation of +0.43 dex (a factor higher than 2.5) in M22 (see Table \ref{t:atmos} Column 8, [Fe/H]$_{IR}$).
Within the uncertaintites, this metallicity spread is in very good agreement 
with previous values based on lower resolution optical data 
\citep[e.g.][]{pila82}, and it is $\sim$ 0.10 dex higher than the value found
by \citet{marino09} using high resolution optical spectra. 
In Fig.\ref{f:fe_comp} we compare our metallicities to those of
\citet{marino09}. For the nine stars in commom, we find that our mean [Fe/H] is higher by +0.13 $\pm$ 0.02 ($\sigma$ = 0.05). 
Interestingly, while \citet{marino11} argue that the spread in [Fe/H] is lower than
that observed in [Ca/H], our data show that the spread in [Fe/H]
(+0.43 dex) is actually about the same as the spread in [Ca/H] (+0.47 dex)
obtained in the optical. 

We note from Table \ref{t:errors} that the total
error in [Fe/H] is about 0.04 dex, while \citet{marino11} estimate a total error
in [Fe/H] of about 0.10 $\pm$ 0.02 dex using optical FeI 
lines (see their Table 4). 
This difference is due to the weaker
sensitivity to T$_{\rm eff}$ of FeI lines with high excitation potential in the
infrared. Indeed, the optical spectroscopic T$_{\rm eff}$ is mainly determined by the
sensitivity to T$_{\rm eff}$ of FeI lines with low excitation potential. 
Even though there are only a few FeI lines in the H band, they are less
sensitive to variations in the stellar parameters.
We thus confirm that there is an intrinsic metallicity
dispersion in M22 as high as $\sim$
0.4 dex. Such a large [Fe/H]-dispersion is not seen in mono-metallic GCs but it is still
smaller than the dispersion found in the complex GC $\omega$ Cen
\citep[see, e.g.][for a detailed discussion]{dacosta09}. 

In addition, \citet{dacosta09} obtained intermediate resolution
spectra at the Ca II triplet for 41 M22 member red giant
stars. They found that the observed M22 [Fe/H]-distribution peaks at 
[Fe/H] = $-$1.90 dex, with their highest abundance star presenting [Fe/H] =
--1.45. Although we have no stars in common with \citet{dacosta09}, we note
that our recovered metallicities agree very well with the observed M22
[Fe/H]-distribution presented in \citet{dacosta09}.

For M22, \citet{bw92} were the first to show that its CN-weak, CN-strong
dichotomy was also accompanied by a dichotomy in [Fe/H] and [$s$/Fe] (see, e.g.,
their Table 5). In their analysis, the CN-weak group has $<$[Fe/H]$>$
= --1.76 and $<$[Ba/Fe]$>$ = 0.05, whereas the CN-strong group shows $<$[Fe/H]$>$
= --1.57 and $<$[Ba/Fe]$>$ = 0.48. These results were recently confirmed by
\citet{marino11} using a larger high quality sample of stars. In Table
\ref{t:abund} we list the final abundances derived in
this work along with the mean [$s$/Fe] for each individual star,
where $s$ stands for Ba and La taken from \citet{marino11}. 
Using our [Fe/H]$_{\rm IR}$ values, we find that the [$s$/Fe]-poor group has $<$[Fe/H]$>$ = --1.72
$\pm$ 0.04 ($\sigma$ = 0.11, N = 6 stars), whereas the [$s$/Fe]-rich group presents $<$[Fe/H]$>$ =
--1.53 $\pm$ 0.05 ($\sigma$ = 0.09, N = 3 stars). 
We conclude that our results are in agreement with those presented in
\citet{kayser08} who found that the number of CN-weak stars in M22 exceeds
the number of CN-strong ones. Also, even though we are dealing with a
limited sample of stars, the results presented in Fig.\ref{f:fe_comp} suggest
that both $s$-poor and $s$-rich groups overlap in [Fe/H]. 

\subsection{Carbon, nitrogen, oxygen and sodium}

\begin{figure}
   \centering
   \includegraphics[width=8.5cm]{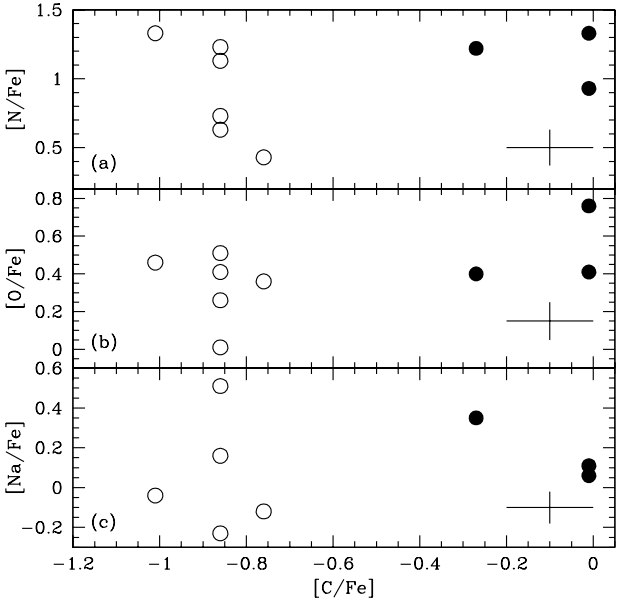}
   \caption{[N/Fe], [O/Fe], and [Na/Fe] as a function of [C/Fe] for the $s$-poor ({\it open
   circles}) and $s$-rich ({\it filled circles}) stars. Typical uncertainties
   are shown.}
              \label{f:cnna}%
    \end{figure}

In Fig. \ref{f:cnna} the [(N,O,Na)/Fe] abundance ratios are plotted as a
function of [C/Fe]. For all stars analysed, the [C/Fe] abundance ratios range from
$-$1.01 to --0.01 while [N/Fe] varies from +0.43 to 1.33.
As shown in Fig. \ref{f:cnna}a, the $s$-poor stars (open circles), with [C/Fe] lower
than --0.7, present a spread in [N/Fe] of 0.8, which is accompanied
by a small variation ($\sim$ 0.3 dex) in [C/Fe]. On the other hand, the
$s$-rich stars (filled circles), with [C/Fe] higher than --0.3, present slightly the
same variation in [C/Fe] ($\sim$ 0.3 dex) that is accompanied by a smaller spread in [N/Fe]
($\sim$ 0.4). These outcomes are in good agreement with previous results
\citep[see, e.g.,][]{brown90,marino11}.

For oxygen, we also estimate a large abundance spread that ranges from 
[O/Fe] = 0.01 to 0.76. 
These values are in good agreement ($\lesssim$ 1$\sigma$) with the
analysis performed in the optical by \citet{marino11} using the forbidden [OI]
line at 630 nm. However, the star III-3 in our analysis presents [O/Fe] that is
higher by 0.35 dex than the mean value found by
\citet{marino11}, $<$[O/Fe]$>$ = 0.41 $\pm$ 0.15. We do not seek to
understand the origin of this difference. Interestingly, M22 follows the
same high [O/Fe] dispersion presented by $\omega$ Cen at the same [Fe/H] range
\citep[e.g.][]{jp10}. Alternatively, \citet{brown90} found a spread of +0.10
dex in [O/Fe] when an
extra reddening is applied to the sample; however, this spread increases to +0.30 dex if
a uniform reddening is assumed. Thus, except for one star, we find that
the $\alpha$-element O is overabundant relative to Fe with values similar to those
seen in halo metal-poor stars. These O overabundances indicate a rapid
chemical evolution history dominated by Type II Supernovae.

Within the limited sample and taking into account the uncertainty of the
measurements, Figs. \ref{f:cnna}a,b show that even though some s-poor giants have
[N/Fe] and [O/Fe] abundances as high as those of at least two of the s-rich
giants, the average [(N,O)/Fe] abundances
in the s-poor M22 stars are lower than in the s-rich group. Sodium, however,
spans a larger amplitude abundance in the
$s$-poor stars compared to the  $s$-rich group (see Fig. \ref{f:cnna}c).

We find a mean $<$A(C+N+O)$>$ = 7.68
$\pm$ 0.25 (N = 9 stars), with a large amplitude abundance of $\sim$ 0.7 dex
($\sim$ 3$\sigma$). From the optical
analysis, \citet{marino11} found $<$A(C+N+O)$>$  = 7.68 $\pm$ 0.16 (N = 14 stars) for
this GC, whereas \citet{bw92} found $<$A(C+N+O)$>$  = 8.04 $\pm$ 0.24 (N = 7
stars). If we consider the $s$-rich and $s$-poor group separately, we find
$<$A(C+N+O)$>$  = 7.93 $\pm$ 0.08 ($\sigma$ = 0.13, N = 3 stars) and
$<$A(C+N+O)$>$  = 7.55 $\pm$ 0.08 ($\sigma$ = 0.19, N = 6
stars) for the $s$-rich and $s$-poor groups, respectively.
From the optical analysis, \citet{marino11} estimated that
the $s$-poor group star has $<$A(C+N+O)$>$ = 7.57 $\pm$ 0.03 ($\sigma$ =
0.09), whereas the $s$-rich group displays $<$A(C+N+O)$>$ = 7.84 $\pm$ 0.03
($\sigma$ = 0.07). These results imply that even though M22 presents a large
amplitude variation in its total A(C+N+O) abundance, within of each M22
$s$-process group the sum A(C+N+O) is constant.  
For our sample, $<$A(C+N+O)$>_{\rm s-poor}$ is smaller than $<$A(C+N+O)$>_{\rm s-rich}$ by $\sim$ 0.40
dex, which confirms the results presented by \citet{marino11}.

For mono-metallic globular clusters that were also analysed in the
infrared, \citet{smith05} found $<$A(C+N+O)$>$ = 8.16 $\pm$ 0.08 for M4, while
\citet{yong08} found $<$A(C+N+O)$>$ = 8.39 $\pm$ 0.14 for NGC 6712. Both GCs
present abundance amplitude of A(C+N+O) lower than 0.35 dex. 
Alternatively, for NGC~1851, another GC displaying multiple stellar
populations, a large 0.6 dex range in A(C+N+O) was also found by
\citet{yong09} analysing high quality optical data of four giant stars. 
However, \citet{villa10} recently claimed to find no significant A(C+N+O) variation in
this GC. While the observed discrepancies in the CNO abundance pattern of NGC
1851 need to be understood,
theoretical models \citep[e.g.][]{cassisi08} suggested that the total A(C+N+O)
abundance in NGC 1851 should be increased by a factor of 2 to better understand the multiple stellar populations. A(C+N+O) also varies
among the different
[Fe/H] stellar groups of $\omega$ Cen \citep[see, e.g.,][]{marino11b}. 
The large abundance spreads we recover for C, N and O might be intrinsically related to the
complex nucleosynthetic history of M22. 

Using Hubble Space Telescope images, \citet{piotto09} 
found evidence for a double sub-giant branch (SGB) for M22 which requires a 
complex star formation history.
While the metallicity spread itself would not explain these two different stellar
populations, the M22's large A(C+N+O) abundance
spread we recover on the RGB could play a key role if it is also acting on the
SGB. We recall that at least for NGC~1851, abundance variations in A(C+N+O)
have been advocated to explain the different stellar populations
\citep{cassisi08}.

\begin{figure}
   \centering
   \includegraphics[width=8.5cm]{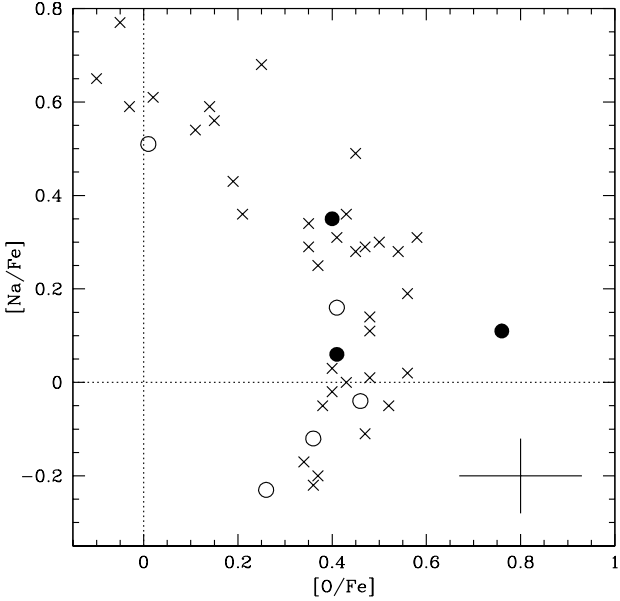}
    \caption{[O/Fe] vs. [Na/Fe] from this study ({\it open and closed circles}) and
    from \citet{marino11} ({\it crosses}). Open and closed circles represent
    $s$-poor and $s$-rich stars, respectively. Typical error bars are indicated.}
              \label{f:oxna}%
    \end{figure}

In Fig. \ref{f:oxna} we plot [O/Fe]
against [Na/Fe] for eight out of nine stars. 
For comparison, we also added in this figure the abundances obtained in the optical
by \citet{marino11} for a larger sample. As seen in this figure, 
both s-rich and s-poor stars in M22 present mean [O/Fe] around 0.40 dex
accompanied by a high Na variation (--0.23 $\leq$ [Na/Fe] $\leq$
0.51). In line with the results found by \citet{marino11}, while our three
s-rich stars present [Na/Fe] $>$ 0, the five s-poor stars range from low ([Na/Fe]
$<$ 0) to high (0.10 $<$ [Na/Fe] $\leq$ 0.51) Na abundances. 
As noted by the referee, the apparent turnover in Fig. \ref{f:oxna} (see also Fig.
14 of \citet{marino11}) is possibly due to the small sample statistics in the case of s-poor giant stars with
[Na/Fe] $<$ 0. At such low Na abundances the Na-O relation seems to become 
very steep and, consequently, there is a substantial change 
in [Na/Fe] over a small change in oxygen around [O/Fe] $\sim$ 0.40 dex. The
Na abundance scatter in such a near-vertical Na-O relation might
artificially produce the visual impression of a correlation between Na-O at
[Na/Fe] $<$ 0. Thus, our independent
analysis at infrared wavelengths also
reveals the well-known Na-O anticorrelation operating in Galactic GCs, which
originates from the enrichment of high-temperature H-burning processed material
from the CNO cycle and NeNa chain.

\subsection{Fluorine}

To date, F has only been investigated in a small 
number of stars in three GCs: $\omega$ Cen 
\citep[][2 stars]{cunha03}, M4 \citep[][7 stars]{smith05} and NGC~6712
\citep[][5 stars]{yong08}. 
For M4 and NGC~6712, the GCs with more than two stars 
with F measurements, the data reveal that fluorine presents 
the largest star-to-star abundance amplitude of all elements and 
that it is correlated with other light elements. 
Therefore, understanding the behavior of F is critical for 
understanding the chemical abundance anomalies in globular clusters.

For M22, we obtained F abundances for seven out of nine stars. A(F)
abundances range from 2.82 to 3.12, which correspond to --0.20 $\leq$
[F/Fe] $\leq$ +0.28 or --0.86 $\leq$ [F/O] $\leq$ +0.27. 
We stress, however, that for stars III-14 and
III-12 we have obtained only the upper limits of their F
abundances due to the weak features of the HF lines (see Figs.
\ref{f:fit5} and \ref{f:fit6}). In contrast to what is seen in M4 and NGC~6712, the amplitude of the A(O) variation
in M22 is {slightly larger} than that of A(F).

As noted, in M4 and NGC 6712 the abundance of F was correlated with the
abundance of other light elements, C, N, O, and Na, and these correlations were
significant. For M22, when we compare the A(F) abundance with A(C, N, O, Na), we
do not find any significant correlations. Therefore, despite the fact that there
is a large F abundance variation in M22, unlike M4 and NGC 6712 there are no
significant correlations between F and the other light elements. The absence of
such correlations may be attributed to the small sample size and/or the more
complex chemical enrichment history of M22 relative to the mono-metallic
clusters M4 and NGC 6712.

\citet{yong08} noted that globular clusters and field stars
seem to define different trends in the A(F)- and [F/O]-A(O) planes, which is in
agreement with the general finding that
globular clusters have abundance patterns that are distinct from those of
the field \citep[see][for a review]{gratton04}. 
A given globular cluster contains stars with O abundances spanning a 
large range of values. The upper envelope of these values is in accord 
with field stars at the same metallicity while the lower envelope extends 
down to stars whose O has been depleted by a factor of 10 or more. 
Similarly, the F abundances show star-to-star variations 
in which the F and O abundances are correlated 
such that the most F-rich objects are also the most O-rich. 
Thus, in the limited data, \citet{yong08} showed that the [F/O] ratio is
constant in a given GC. They detected a linear decrease of A(F) with A(O),
accompanied by a constant [F/O] ratio as the A(O) abundances decrease, which
suggests that 
the nucleosynthetic process(es) responsible for the light element 
abundance variations deplete F and O by the same amounts. If this is the case,
the [F/O] discrepancy between  field stars and GCs is likely driven by
unusually low F abundances in GCs relative to field stars. 

\begin{figure}
   \centering
   \includegraphics[width=8.5cm]{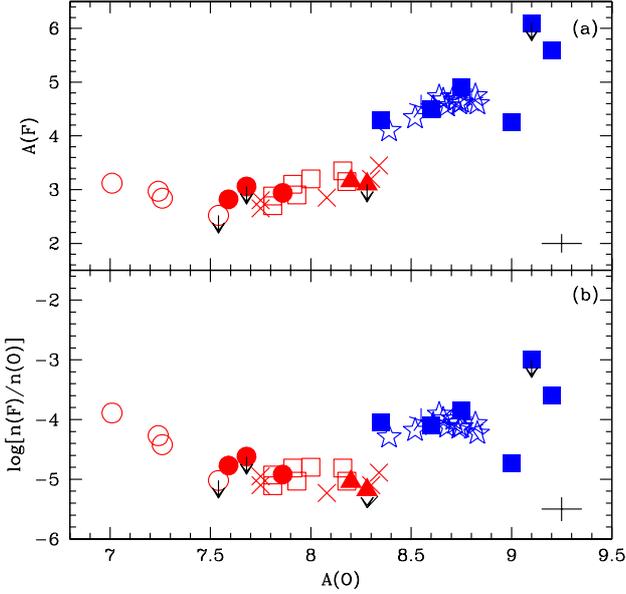}
   \caption{(a) A(F) vs. A(O) and (b) log[n(F)/n(O)] vs. A(O).
   Symbols stand for: $s$-rich M22 stars ({\it filled circles}),
   $s$-poor M22 stars ({\it open circles}), NGC~6712 ({\it
   crosses}: Yong et al. 2008), M4 ({\it open squares}: Smith et al. 2005), $\omega$
   Cen ({\it filled triangles}: Cunha et al. 2003), bulge stars ({\it filled squares}: Cunha et al. 2008), and
   field stars ({\it stars}: Cunha et al. 2003; Cunha \& Smith 2005).}
              \label{f:fluor}%
    \end{figure}

Following \citet{yong08}, we show in Fig. \ref{f:fluor} the abundances of A(F)
and log[n(F)/n(O)] against A(O) for M22 compared with those values in different stellar components
and environments (field and GCs), as indicated in the figure's caption. 
Looking at the M22 sample as a whole, the main conclusion is that A(F)
remains nearly constant with A(O), with $<$A(F)$>$ = 2.89 $\pm$ 0.07 ($\sigma$ =
0.20 dex, N = 7 stars), whereas
the F/O abundance drastically
increases for A(O) abundances lower than 7.5 dex. Rejecting the upper limits on F, the average A(F) for the $s$-rich stars
is $<$A(F)$>$ = 2.88 $\pm$ 0.06 ($\sigma$ =
0.08 dex, N = 2 stars), while for the $s$-poor stars $<$A(F)$>$ = 2.98 $\pm$
0.08 ($\sigma$ = 0.14 dex, N = 3 stars). Within the
limited data, the $s$-poor sample has a larger F
abundance dispersion than the $s$-rich sample. However, within the uncertainties, the two
populations have the same mean A(F). 

On the other hand, a detailed inspection of Fig. \ref{f:fluor} suggests that
for our limited number of stars in M22, the sample is well divided at A(O) =
7.5 dex. In this figure, the stars III-15, IV-97 and IV-102 have the lowest A(O)
abundances or [O/Fe] = (+0.01, +0.36, +0.41), which are in good
agreement with those found by \citet{marino11}, that is, [O/Fe] = (+0.11,
+0.40, +0.43), respectively. In addition, as seen in Figs. \ref{f:fit4},
\ref{f:fit5} and \ref{f:fit6}, the HF features in these three stars are relatively strong,
implying that their F abundances are not upper limits.
Thus, whether one subdivides the sample based on their A(O) metallicities, four
M22 member giant stars (3 $s$-rich and 1 $s$-poor) follow the general linear
A(F)-A(O) trend found by \citet{yong08} for three Galactic GCs, albeit with a
higher dispersion. Additional F measurements of more metal-poor field and GC
stars are required in order to draw firm conclusions about the Galactic
A(F)-A(O) at low metallicities. 

\begin{figure}
   \centering
   \includegraphics[width=8.5cm]{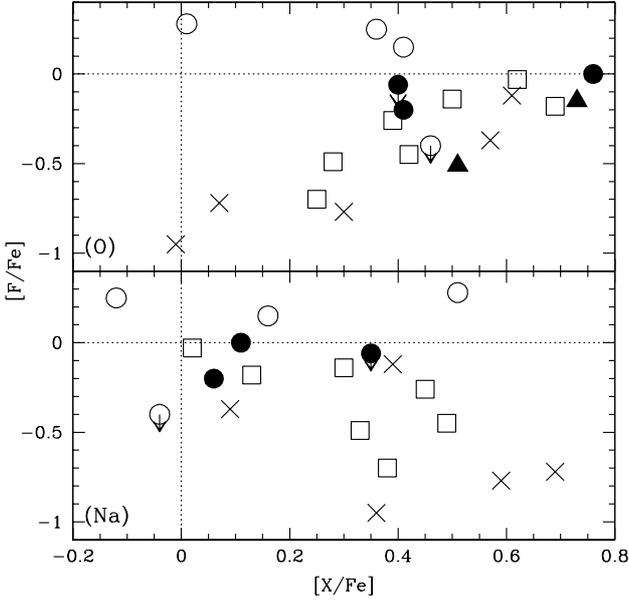}
   \caption{[F/Fe] against [O/Fe] ({\it top}) and [Na/Fe] ({\it bottom}).
   Symbols are as given in Fig. \ref{f:fluor}.}
              \label{f:fluorfe}%
    \end{figure}

In Fig. \ref{f:fluorfe} we show the F and Na-O abundance diagram in M22 compared to that
seen in $\omega$ Cen, M4 and NGC~6712. Overall, there is a general trend for
[F/Fe] to increase with [O/Fe] and decrease with [Na/Fe] in the different GCs. 
Once again, an
intriguing feature of this figure concerns the location of the three stars
with the lowest A(O) abundances in M22. Within the
observed uncertainties, the stars IV-97 and
IV-102 do follow the global [F/Fe]-[O/Fe]:[Na/Fe] trend at [O/Fe] $\sim$ 0.40.
However, the most F-rich ([F/Fe] = +0.28), O-depleted
([O/Fe] = +0.01), and Na-rich ([Na/Fe] = +0.51) star in M22, III-15, clearly
does not follow this general trend. Instead, the observed [F/Fe] abundance ratio
of the star III-15 differs by more than 1 dex with respect to what
is observed for NGC~6712 at nearly the same [O/Fe] ratio. This result is very
intriguing with no easy explanation as III-15 is confirmed
kinematically and spectroscopically as a M22 member.

As discussed in \citet{kobayashi11} both the $\nu$-process of core-collapse
supernovae (SNe II and HNe) and AGB stars play a crucial role in the production
of F. Nevertheless, the relative contribution from low-mass supernovae may
be smaller in the GCs than in the field. 
F production in AGB stars is dominated by the contribution of
stars with masses $\approx 1-3 M_{\odot}$, depending on
metallicity \citep[see also the yields of][]{karakas10a}.
For these stars, the abundances of C and F need to be considered
together with the $s$-process elements because the F production in AGB stars
occurs in the He-intershell via a complex series of proton, $\alpha$, and 
neutron-capture reactions, beginning with the  
$^{14}$N($\alpha$,$\gamma$)$^{18}$F($\beta$$^{+}$)$^{18}$O({\it
p},$\alpha$)$^{15}$N($\alpha$,$\gamma$)$^{19}$F reaction. 
The protons for the CNO cycle reaction $^{18}$O($p,\alpha$)$^{15}$N
come from the $^{14}$N($n,p$)$^{14}$C reaction, which in turn requires 
free neutrons. The dominant neutron source in low-mass AGB stars for
F and $s$-process element production is the
$^{13}$C($\alpha$,n)O$^{16}$ reaction \citep[e.g.,][]{busso01}.
Furthermore, because F production occurs in the He-intershell, it is dredged to the
surface via the repeated action of the third dredge-up (TDU), along
with C and any $s$-process elements. 

It is important to remember that the lifetimes of AGB stars with
$M \lesssim 3 M_{\odot}$ are relatively long ($\tau \gtrsim 300$Myr).
It is not clear if these stars will have had time to contribute toward
the bulk chemical enrichment of a forming GC, which show no
discernible age spread \citep[see e.g.][]{mf09}. It is partly for this reason that the
contribution of intermediate-mass AGB stars, with masses between
$4-8 M_{\odot}$, is so attractive.  In intermediate-mass AGB stars, 
the base of the convective envelope can reach the top
of the H-burning shell, causing proton-capture nucleosynthesis 
to occur there (known as hot bottom burning, HBB). The main result
is that CNO nuclei are converted into N, and F is efficiently
destroyed. One of the biggest uncertainties to effect models of
intermediate-mass AGB stars is the unknown efficiency of the TDU.
At low metallicity, theoretical models predict either no or little 
TDU \citep{ventura09a} or predict that intermediate-mass AGB stars
experience efficient TDU \citep{herwig04b,karakas10a}. Models
with efficient TDU would also produce $s$-process elements via
the $^{22}$Ne($\alpha$,n)Mg$^{25}$ neutron source that operates
during convective thermal pulses. The main signature would be
copious production of elements near the first $s$-process peak
(e.g., Rb, Sr, Y, Zr) and much smaller quantities of Ba and Pb
\citep[e.g., the metallicity Fe/H = --2.3 models of][]{lugaro11}. 
One implication of an efficient TDU is that the total number of
C+N+O nuclei is not conserved.

\begin{figure}
   \centering
   \includegraphics[width=8.5cm]{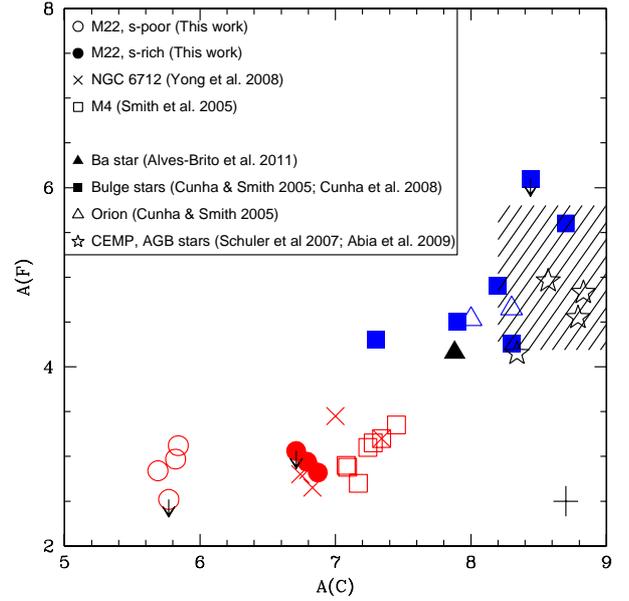}
   \caption{A(F) vs. A(C) for different objects in the Galaxy.
   Symbols stand for: $s$-rich M22 stars ({\it filled circles}),
   $s$-poor M22 stars ({\it open circles}), NGC~6712 ({\it
   crosses}: Yong et al. 2008), M4 ({\it open squares}: Smith et al. 2005), bulge stars ({\it filled
   squares}: Cunha et al. 2008), Ba star ({\it filled triangle}: Alves-Brito et
   al. 2011 ), Orion
   giant stars ({\it open triangles}: Cunha \& Smith 2005; Cunha et al. 2008), CEMP and AGB stars
   ({\it stars}: Schuler et al. 2007; Abia et al. 2009). The hachured area
   indicates the A(F)-A(C) position of Galactic AGB stars analysed in Abia et
   al. 2010.}
              \label{f:fluorcarbon}%
    \end{figure}

In Fig. \ref{f:fluorcarbon}, which shows A(F) as a function of A(C), we find further evidence for the chemical
dissimilarity between F abundances in the field from those of GCs.
Interestingly, all field giant stars analysed (regardless of their [C/Fe]-enhancements)
have higher A(F) abundances than those of GCs. We note, however, that there is
no overlap in [Fe/H] between the field and GC stars.  
Within the limited sample, the M22 C- and s-rich
population shares a similar A(F)-A(C) relation compared to the stars analysed in
other  
GCs. On the other hand, the M22 C- and s-poor population does not fit the
general linear A(F)-A(C) trend seen for the other GCs.
We would like to reinforce once again that the
three comparison GCs for which F data are available are all more oxygen-rich
than the A(O)-poorest giants in the M22 
sample. Hence, the significance of the non-linearity in the M22 A(O)-A(F) data at 
low abundances is still unclear. Additional F data for other low-metallicity GCs are
required to determine whether the A(O)-A(F) relation in M22 is unusual as a
result of its complex nucleosynthetic history or, alternatively, whether the GC 
population in general shows a non-linear A(O)-A(F) relation at low [Fe/H].

\begin{figure}
   \centering
   \includegraphics[width=8.5cm]{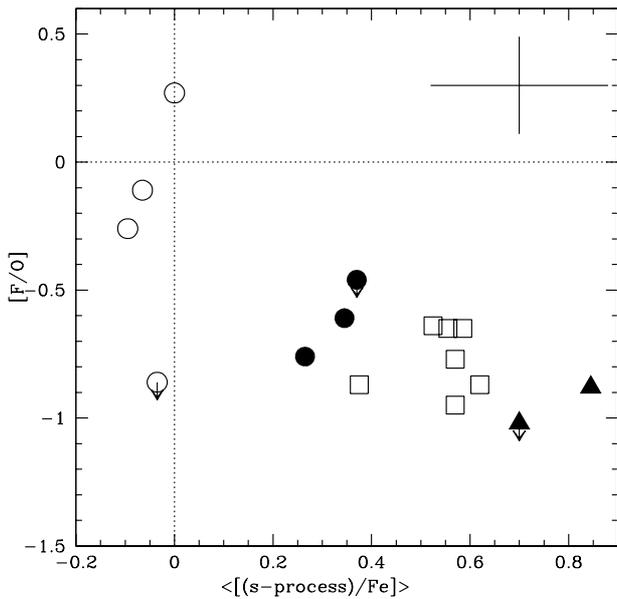}
   \caption{[F/O] vs. $<$[($s$-process)/Fe]$>$ for $s$-rich M22 stars ({\it
   filled circles}), $s$-poor M22 stars ({\it open circles}), M4 ({\it open
   squares}: Smith et al. 2005), and $\omega$
   Cen ({\it filled triangles}: Cunha et al. 2003).}
              \label{f:fluoroxpesados}%
    \end{figure}

Figure \ref{f:fluoroxpesados} displays [F/O] against the mean
[($s$-process)/Fe]. From this figure, there is evidence for an
anticorrelation
between the [F/O] and $s$-process enhancements in GCs in the sense that $s$-rich stars
also have lower [F/O] abundances.
In the solar system, the main component of the $s$-process contributes to
about 84\% of the Ba abundances, whereas $\approx$16\% is
 due to the $r$-process \citep[see, e.g.][]{trav99}; for La, this proportion is of
 61\% ($s$) and 39\% ($r$).
 While the $r$-process is related to the final stages
 of evolution of massive stars (M $>$ 8 M$\sb \odot$), the $s$-process main
 component is believed to occur in AGB stars of low (1-3 M$\sb \odot$) 
 or intermediate (4-8 M$\sb \odot$) masses \citep[see, e.g.][]{busso99}. 
However, as noted above, the production of 
$s$-process elements in intermediate-mass AGB stars is
accompanied by F destruction and could account for the anti-correlation
seen in Fig. \ref{f:fluoroxpesados}. Theoretical predictions of
$s$-process nucleosynthesis from intermediate-mass AGB stars at the
metallicities of M22 are needed to verify this scenario. 

  For NGC~6712, \citet{yong08} found that its [F/O]
  abundance amplitude is compatible with a production in massive (M$\gtrsim$
  5M$_{\odot}$) AGB stars. Unfortunately, there are no observational clues about the $s$-process
  enrichment in NGC~6712. For M4, not only its high $s$-process enrichment
  \citep{ivans99} but also its [F/O] abundance ratio \citep{smith05} suggest a chemical
  enrichment history dominated by intermediate AGB stars.

M22 and $\omega$ Cen, however, present discrete stellar populations in their
colour-magnitude diagrams, which means they have experienced a more complex
nucleosynthetic history.
For $\omega$ Cen, \citet{cunha03} argue that the low [F/O]
abundances obtained for two $s$-rich stars cannot be
explained by the production of F as the result of low-mass metal-poor AGB stars.
M22, on the one hand, presents a sharp separation between its $s$-rich and
$s$-poor groups, which can be explained in terms of its double stellar
populations \citep{marino09}.  
After testing different chemical 
models to explain the observed [F/O] abundances in all GCs studied to 
date (including our preliminary results), \citet{kobayashi11} showed 
that the observed [F/O] abundances in GCs are consistent with models 
that include the AGB yields only. However, given the timescale
considerations noted above, an alternative scenario might be to
consider a small contribution from SNe in GC (these produce F via the
$\nu$ process) followed by the destruction of F via HBB in
intermediate-mass AGB stars. These AGB stars have also produced
$s$-process elements along with large amounts of helium \citep[see
discussion in][]{roederer11}. \citet{n04} claims that massive stars could be the source of the high helium
necessary to explain the different stellar populations on the main
sequence of $\omega$ Cen. As M22 shares chemical properties with $\omega$ Cen,
this explanation could also, in principle, be extended to M22.

\section{Conclusions}
 
We have acquired high-resolution infrared data for a sample of nine stars in the
Galactic globular cluster M22. The main goal was to obtain high precision
abundances of C, N, O, F, Na and Fe using atomic and molecular lines
in the H and K bands. 

We confirm with high quality data at infrared wavelengths
that M22 resembles the chemical properties seen not only in mono-metallic GCs
but also in other massive GCs displaying multiple stellar populations. We have
found/confirmed that:

\begin{itemize}

 \item there is a spread in [Fe/H] in M22 of $\sim$ 0.4 dex; 
 \item there is a large dispersion in the light C, N, and O elements. Although
 A(C+N+O) spans a large amplitude abundance ($\sim$0.7 dex), it is
 approximately constant within each M22 $s$-process group; 
 \item O and Na are anticorrelated;
 \item the abundance spread for F is $\Delta$A(F) = 0.6 dex, an amplitude
 comparable to that seen for other light elements in M22;
 \item while the four most A(O)-rich stars are
consistent with the general linear A(F)-A(O) trend from other Galactic GCs, the three
most A(O)-poor stars in M22 do not follow this trend. More data at low
metallicity are required.

\end{itemize}
The overall abundance pattern of M22 suggests that this GC might have
experienced a complex star formation history, where the large abundance spread
is likely primordial. In this framework, 
intermediate-mass AGB stars
might have polluted the medium where these stars were formed. Further
investigation is necessary to check if these large abundances
variations are also found on the SGB of M22 or even at an earlier phase of
stellar evolution.

\begin{acknowledgements}
A.A.B. acknowledges support from FONDECYT-Chile (3100013), CNPq-Brazil (PDE,
200227/2008-4), and ARC (Super Science Fellowship, FS110200016).
J.M. thanks support from FAPESP (2010/50930-6), 
USP (Novos Docentes), and CNPq (Bolsa de produtividade). 
We wish to thank Dr. Stuart Ryder, from the
Australian Gemini Office, for his
assistance during the Phase II definitions of our Gemini/Phoenix observations,
and also Dr. Javier Alonso-Garcia for providing us Fig.
\ref{f:map_extinct}. We warmly thank Prof Gary da Costa for reading the
manuscript and for his helpful comments and remarks. We also thank the
anonymous referee for helping improving the paper. 
Based on observations obtained at the Gemini Observatory, which is operated by
the AURA, Inc., under a cooperative agreement with the NSF on behalf of the
Geminipartnership: the NSF (United States), the STFC (United Kingdom), the NRC
(Canada), CONICYT (Chile), the ARC (Australia), CNPq (Brazil) and SECYT
(Argentina). This paper uses data obtained with the Phoenix infrared
spectrograph, developed and operated by the National Optical Astronomy
Observatory. The spectra were obtained as part of the program GS-2009A-Q-26.
This publication makes use of data products from the Two Mass Micron All Sky
Survey, which is a joint project of the University of Massachusetts and the
Infrared Processing and Analysis Center/California Institute of Technology,
funded by the National Aeronautics and Space Administration and the National
Science Foundation. 
\end{acknowledgements}


\bibliography{}

\end{document}